# Mapping Complex Technologies via Science-Technology Linkages; The Case of Neuroscience

### – A transformer based keyword extraction approach –

Daniel S. Hain[φ], Roman Jurowetzki[φ], and Mariagrazia Squicciarini[†]

[φ]*AI:Growth Lab, Aalborg University Business School, DK*
[†]*UNESCO, Social and Human Sciences Sector, FR*

May 19, 2022

**Abstract:** In this paper, we present an efficient deep learning based approach to extract technology-related topics and keywords within scientific literature, and identify corresponding technologies within patent applications. Specifically, we utilize transformer based language models, tailored for use with scientific text, to detect coherent topics over time and describe these by relevant keywords that are automatically extracted from a large text corpus. We identify these keywords using Named Entity Recognition, distinguishing between those describing methods, applications and other scientific terminology. We create a large amount of search queries based on combinations of method- and application-keywords, which we use to conduct semantic search and identify related patents. By doing so, we aim at contributing to the growing body of research on text-based technology mapping and forecasting that leverages latest advances in natural language processing and deep learning. We are able to map technologies identified in scientific literature to patent applications, thereby providing an empirical foundation for the study of science-technology linkages. We illustrate the workflow as well as results obtained by mapping publications within the field of neuroscience to related patent applications.

**Keywords:** Patent landscaping, technology impact assessment, deep learning, natural language processing

## 1. Introduction

Times of accelerating technological change (Butler, 2016), growing interdisciplinarity of science and technology (Porter and Rafols, 2009) as well as increasing complexity of science and technology systems (Catalán et al., 2020) call for fast and responsive techniques to quantify, map, and understand technology development. Doing so is paramount to assess, influence, and facilitate the development of complex and interdisciplinary emerging technologies with potentially high economic and social impact, such as neuroscience, genetics, and artificial intelligence.

A large share of technology development builds on advances in science, where existing studies have demonstrated the rising interactions between science and technology and the important role science has played in accelerating technologies (Meyer-Krahmer and Schmoch, 1998; Acosta and Coronado, 2003). Consequently, current developments in science are important signals for identifying future's promising technologies.

However, measuring and mapping of science-technology linkages has been proven challenging (Mansfield, 1991; McMillan et al., 2000; Narin et al., 1997). There are several data sources that are leveraged to identify direct science-technology linkages (Bekkers and Freitas, 2008), such as collaborations (Giunta et al., 2016) or the citation of non-patent-literature (NPL) in patents (Acosta and Coronado, 2003; Narin et al., 1997). However, such direct measures tend to be sparse and biased, since for instance joint university-industry patenting is the exception rather than the norm, and NPL citations are rarely used. Without such direct traces, author-matching and natural language processing (NLP) techniques have been used to identify paper-patent pairs (e.g., Magerman et al., 2015) which both relate to the same invention. Again, due to division of labor and corporate IP strategies it is not uncommon that publication authors are not the same as the named inventors on a related patents, leading to similarly sparse results.

Alternative indirect approaches to identifying science-technology linkages and to map technology applications arising from a particular field of scientific research focus on the identification of technologies within scientific literature, and in a separate step map them to technologically similar patents.

In this paper, we present a deep learning (DL) based approach to extract technology-related topics within scientific literature, and identify these technologies within patent applications. In detail, we utilize named entity recognition (NER) techniques and transformer-based text embeddings which are specifically fine-tuned to scientific literature to extract technology keywords. In a next step we cluster these keywords into human interpretable topics describing a certain technology application by a set of keywords related to the method as well as application and issues addressed. Finally, we create a large amount of search queries based on combinations of methods and applications, which we use in a semantic search to identify related patents.

We illustrate the workflow as well as results obtained by mapping publications within the field of neuroscience to related patent applications. This represents an interesting field to demonstrate the advantages of our approach, since neuroscience research has traditionally been an interdisciplinary research field where scientists from different disciplines like medicine, psychiatry, psychology, biology, biochemistry and linguistics work together (Schwechheimer and Winterhager, 2001). Throughout the last decade, technological change as well as changing global needs have lead to scientific as well as technological paradigm shifts in neuroscience and leading to development of the notion of neurotechnology. Research strands such as brain imaging, brain connectivity (Yeung et al., 2017), and lately neural computing (Savage, 2019) and neurodevices such as Brain Computer Interfaces (BCIs) have added to the diversity of technologies and potential applications. This degree of interdisciplinarity and dynamism limits the usefulness of traditional approaches to technology mapping via existing patent classifications. This has resulted in a limited overview over the types of technology applications in neuroscience, their scale, impact, and geography. Recently, academia and policymakers alike have recognised potential ethical and legal concerns related to the potentials of the latest generation of neurotechnology (Allhoff et al., 2011; Anderson et al., 2012; Drew, 2019; Dubljević et al., 2017; Jarchum, 2019), stipulated the formulation of guiding principles and regulations for future development (Garden et al., 2019). However, the formulation as well as the practical implementation of such principles requires a common understanding of what neuroscience is and is made of, in terms of both scientific and technological developments (neurotechnology), as well as possible applications.

By doing so, we aim at contributing to the growing body of research on text-based technology mapping and forecasting that leverages latest advanced in NLP and DL (Hain et al., 2022b,a, e.g., cf.). More specifically, we provide an approach that identifies technology related topics in scientific literature, and allows to disentangle keywords to identify methods and techniques embodied in the technology from the context and issue it is applied to. In contrast to Subject-Action-Object (SAO) (e.g., Yang et al., 2017) or ontology-based (e.g., Soo et al., 2006) approaches which are labor intensive, our work-flow allows to do so almost fully automated and is thereby highly scalable. We are able to map these technologies identified in scientific literature to patent applications, thereby providing an empirical foundation for the study of indirect science-technology linkages (e.g., Mansfield, 1991; Bekkers and Freitas, 2008; Acosta and Coronado, 2003). In contrast to existing approaches, neither a direct explicit link between publications and patents such as a NLP citation, nor the use of an exactly matching keyword terminology is necessary to identify applications of certain technologies. Our approach uses latest developments within transfer learning (Luan et al., 2018; Cohan et al., 2020; Grootendorst, 2020) in NLP to replicate, automate and scale the process of technology search as it is performed by a domain expert that, for instance, searches the patent landscape for related technologies during a novelty screening. The expert would typically derive derive relevant keywords from the description of a potential invention in part from the present text and in part relying on his or her expertise. While appropriate for individual inventions, such traditional and manual approaches are costly, time-consuming, and thus impractical for exercises at the present scale (Park and Yoon, 2017). Having identified related technologies we provide an inclusive and and interdisciplinary mapping of patent applications related to neuroscience technology, we are able to inform the current discussion on the potential needs for regulatory frameworks targeting these technologies. The results of this mapping can be accessed and explored via an interactive application at `https://neurscience-sci-pat.herokuapp.com/`.

The remainder of this paper is structured as follows. In section 2, we review the state-of-the-art of the literature on science and technology mapping, and the identification of science-technology linkages. In section 3, we discuss methodological considerations and describe our approach to extract technology keywords, cluster them into technology topics, and map these to patent applications. In section 4, we explore the results of our analysis for the case of neuroscience patents and demonstrate potential research applications. Finally, section 5 offers concluding observations and points towards promising avenues for future research.

## 2. Background and Literature

### 2.1. Mapping Science-Technology Linkages

Identifying science-technology linkages has been extensively studied in the past decades, in part to understand technological evolution but also to document impact of (public) investment into research (Meyer, 2000; Looy et al., 2003). Methods for linking science to patented technology can be broadly distinguished into 3 types. (1) Approaches that depart from science-originating metadata that is identified in patents. Here we find early approaches that identify and count patents filed by universities and individual researchers (Henderson et al., 1994; Schmoch, 1997). Related to that, author-inventor matching has been used to establish links between patents and scientific research (Boyack and Klavans, 2008; Cassiman et al., 2007). A typical issue here is the fact that person disambiguation (especially when working with East-Asian names) is challenging in itself and sometimes these studies would focus on rare names. (2) Citation-based approaches have been proposed as another approach, focusing on non-patent literature (NPL references, mostly scientific publications) found in patents (Noyons et al., 1994; Schmoch, 1997; Meyer, 2000; Verbeek et al., 2002; Callaert et al., 2012). While this approach typically provides well documented, explicit linkages (sam-ples), it has two major drawbacks. First, it only works where citations to scientific literature are explicitly made, which more often than not is not the case – only a third of all patents (Callaert et al., 2006) belonging to particular industries (Looy et al., 2003) refers to NPL. Second, it is not a straightforward approach where links need to be established going from science to technology rather than vice versa, i.e. where patents are not known *a priori* and need to be identified departing from scientific publications. One of the few exceptions here is the work by Glänzel and Meyer (2003) that tracks "reverse citations", i.e. patent citations in scientific literature, finding that these are few and concentrated in specific domains – mainly chemistry, pharmaceuticals and medicine. Furthermore, it is important to mention that legal requirements regarding the inclusion of citations in patents differ around the world with USPTO filed patents having consistently higher shares of NPL citations compared to EPO applications (Michel and Bettels, 2001). A critical argument beyond immediate shortcomings of methods using observable explicit connections (such as patents) is that it assumes a linear model of innovation and technological development (Narin et al., 1997). This may be an oversimplified view (Tussen et al., 2000), especially considering dominant conceptual frameworks that describe innovation and emergence of technology, from the (national) innovation system (Nelson, 1993; Lundvall, 1992) to the Triple Helix (Etzkowitz and Leydesdorff, 2000) and more recent ecosystem frameworks (Adner and Kapoor, 2010) that all highlight the importance of non-linearity and interdependence, suggesting that while scientific discovery may lay the foundation and shape trajectories for technological development, explicit traces can be hard to identify.

(3) Lastly, there are "content based" approaches, that the combination of techniques used in this paper also belongs to. Given the proliferation and variety of machine learning methods here we find different and creative combinations of methodologies – sometimes they are mentioned using the notions of "text mining" and "tech mining" (Porter and Cunningham, 2004). There is a large number of projects applying these technologies within either the science or technology domain (e.g., Zhang et al., 2016), but contributions in which cross-domain linkages are established are more seldomly found. Bakhtin et al. (2017) use term co-occurrence to establish linkages between scientific literature and policy documents. Ranaei et al. (2017) use Latent Dirichlet Allocation (LDA) to identify topics within science and patent text simultaneously, thus inferring linkages from co-occurrence within a topic. Shibata et al. (2011) use a combination of citation network analysis and NLP, in particular different measures of semantic similarity, to detect technological frontiers.

## 2.2. Technochnology Mapping and Landscaping

# 3. Data and Methods

## 3.1. Data

### 3.1.1. Publication data

To identify neuroscience related research, we query the Scopus database for all English-language publications in the period 2000 until 2021 which are categorized under the corresponding Neuroscience subject area.[1] Subject areas are assigned on journal level based on the aims and scope of the title, and on the content it publishes. Consequently, false positives (unrelated article published in journal with neuroscience subject area) as well as false negatives (related article published in journal without neuroscience subject area) on publication level are possible.[2]

This resulted in 1,045,623 publications within neuroscience related journals during the period of interest. Figure 1 illustrates the development of annual publications, which increase from slightly above 30,000 in 2000 to almost 80,000 in 2021. Among these publications, we extracted bibliographic data for the 2000 most cited publications per year.[3] We thereby assume the most cited publications to be the most relevant in terms of their impact on the development of future technology, hence are a suitable subset to identify technology-related keywords and topics.

---

[1] We use the search query "SUBJAREA ( neur ) AND PUBYEAR > 1999 AND PUBYEAR < 2022 AND ( LIMIT-TO ( DOCTYPE , "ar" ) OR LIMIT-TO ( DOCTYPE , "cp" ) ) AND ( LIMIT-TO ( LANGUAGE , "english" ) )".

[2] The Scopus Subject area Neuroscience contains the following subjects: Biological Psychiatry, Cellular and Molecular Neuroscience, Cognitive Neuroscience, Developmental Neuroscience, Endocrine and Autonomic Systems, Neurology, Sensory Systems, and Neuroscience (miscellaneous).

[3] This number is a consequence of the Scopus quota limits. Since the number of publications vary over time, a relative selection such as the top 10 percent top cited publications per year would be preferable.

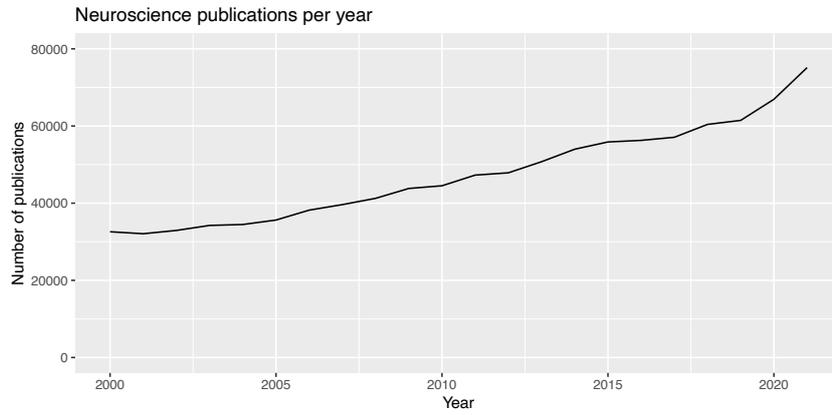

Figure 1: Number of neuroscience publications

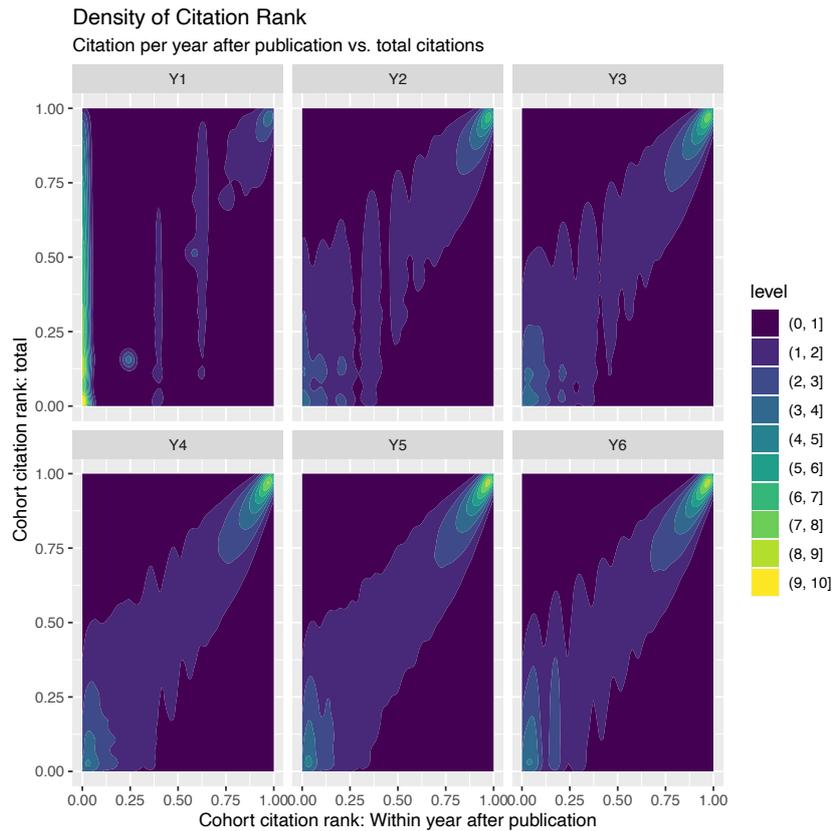

Figure 2: Density of citation by year after publication

However, for publications from the most recent years this information is truncated,

since citation accumulate over time. In Figure 2 we provide a density-plot of the cohort citation rank of a publication received in the corresponding year against its total cohort citation rank.[4] We do so for the year of publication (Y1) until the 5th year afterwards (Y6). We see that the citations received in Y1 only loosely correspond with the paper overall receives. This already changes in Y2, where the association between annual citation rank and total citation rank becomes stronger. This is association is particularly strong for the upper levels of citation ranks, meaning that citations received in Y2 and on wards are a strong signal for the papers overall citations received. In conclusion, selecting the most cited publications within cohort is an efficient strategy to identify high (citation) impact science already a year after publication of the article.

Table 1: Neuroscience Publications, Summary

| Publications | | Journals | | Author Keywords | |
|---|---|---|---|---|---|
| Country | n | Journal | n | Keywords | n |
| USA | 14853 | JOURNAL OF NEUROSCIENCE | 1866 | ALZHEIMER'S DISEASE | 1199 |
| UK | 3326 | NEURON | 1674 | DEPRESSION | 859 |
| GERMANY | 2711 | NATURE NEUROSCIENCE | 1108 | PARKINSON'S DISEASE | 776 |
| CHINA | 1944 | NEUROIMAGE | 981 | HIPPOCAMPUS | 730 |
| CANADA | 1593 | BRAIN RESEARCH | 897 | SCHIZOPHRENIA | 700 |
| JAPAN | 1448 | EMBO JOURNAL | 847 | INFLAMMATION | 686 |
| ITALY | 1366 | ELIFE | 701 | FMRI | 666 |
| FRANCE | 1309 | BIOLOGICAL PSYCHIATRY | 668 | MICROGLIA | 590 |
| NETHERLANDS | 1039 | NEUROSCIENCE LETTERS | 661 | NEURODEGENERATION | 498 |
| AUSTRALIA | 1023 | MOLECULAR PSYCHIATRY | 592 | STRESS | 495 |
| SPAIN | 686 | NATURE REV. NEUROSCIENCE | 553 | COGNITION | 494 |
| SWITZERLAND | 638 | NEUROSCIENCE | 493 | ANXIETY | 468 |
| SWEDEN | 497 | ANNALS OF NEUROLOGY | 483 | AGING | 467 |
| KOREA | 375 | NEUROSCIENCE & BIOBEHAV.L REV. | 474 | DOPAMINE | 454 |
| BRAZIL | 373 | NEUROCOMPUTING | 456 | COVID-19 | 439 |
| BELGIUM | 348 | MOVEMENT DISORDERS | 424 | STROKE | 423 |
| ISRAEL | 347 | PLOS BIOLOGY | 422 | NEUROINFLAMMATION | 422 |
| DENMARK | 292 | TRENDS IN COGNITIVE SCIENCES | 416 | EPILEPSY | 418 |
| AUSTRIA | 259 | EUROP. JOURNAL OF NEUROSCIENCE | 398 | MULTIPLE SCLEROSIS | 415 |
| INDIA | 236 | INVEST. OPHTHALMOLOGY & VISUAL SCIENCE | 363 | META-ANALYSIS | 407 |

*Note.* This table reports most productive countries, top journals, and keywords in neuroscience based on the annually most cited publications in the period 2000-2021.

Table 1 provides a summary over top countries and journals in terms of most cited neuroscience publications, and the top author-assigned keywords used in them. Overall, among the results appear intuitive. We see the USA is leading the field of neuroscience in terms of number of publication by a substantial margin, followed by the UK, Germany and China. In terms of Journals, we among the most represented journals see unsurprisingly many journals dedicated to the broad field of neuroscience (e.g., Journal of Neuroscience, Neuron, Nature Neuroscience). We also notice journals specialized on subfields of neuroscience (e.g., Neuroimage), and journals with broader focus areas which to some extend overlap with neuroscience (eg. PLOS Biology, Neurocomputing, Trends in Cognitive Science). In summary, the corpus appears to capture the broad field of neuroscience research and the represented disciplines well, even though the broad and interdisciplinary nature of some of the included journals is likely to lead to false positives, meaning included publications which are content-wise not or only marginally related to neuroscience.

---

[4]Figure A.1 and Figure A.2 provide additional visualizations of the distribution of citations by year of publication as well as year after publication.

### 3.1.2. Patent data

The patent data we use for our study was retrieved from the EPO's Worldwide Patent Statistical Database (PATSTAT, Autumn 2021 edition), which covers bibliographic patent data from more than 100 patent offices over a period of several decades. We include all patents containing an English language abstract in the period from 2000 to 2020. To avoid duplicates caused by filing the same patent at multiple patent offices, follow De Rassenfosse et al. (2013) and only include priority filings, meaning the first filed application of a patent. In order to focus on technologies and patents with potentially global impact, we only include priority patents from (DOCDB) patent families including at least one application at one of the IP5 patent offices[5]

We further enrich the patent information to be found in PATSTAT with several additional data sources, such as the extended and geocoded information on applicant and inventor location provided by De Rassenfosse et al. (2019), and calculate common indicators of patent quality (cf. Squicciarini et al., 2013) in order to evaluate our results, and map the worldwide development of neurotechnologies.

Within the resulting list of IP5 priority patents, we identify patents likely to be related to neurotechnologies by applying the approach explained in the following subsection. The results are described and discussed in the following section 4.

---

[5]Which includes the United States Patent and Trademark Office (USPTO), the European Patent Office (EPO), the Japan Patent Office (JPO), the Korean Intellectual Property Office (KIPO), and the National Intellectual Property Administration (CNIPA formerly SIPO).

## 3.2. Methods

The core challenge in this study is the identification of neuroscience patents, a broad and emerging field embracing multiple technologies from various disciplines. While we were able to use the broad subject area classification within *Scopus* for the identification of neuroscience research, there are no explicit categories for such interdisciplinary fields for patented technologies. The diversity of technologies and applications as well as rapid technological development prevents a inclusive identification via static technology classifications such as IPC or CPC.

In this paper we rely on a multi-step approach where we start with automated identification of latent themes (or topics) and keywords in the research literature. For each of these themes we use detected keywords and key-phrases to construct search queries that are subsequently used for semantic search within the patent database. This approach, illustrated in Figure 3, allows us to combine a broad scope of the overall search – neuroscience - with being able to restrict results to very specific subfields such as for instance "Sleep and Memory Consolidation". The approach can be combined with domain expertise, where after the detection of topics a domain expert would select those that are deemed relevant as foundation for technology identification. We report overall results for patents matched to all topics as well as selected topics that – without profound domain expertise – can be attributed to neurotechnology.

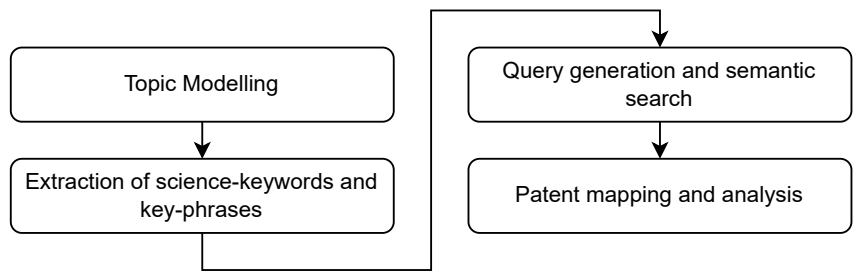

Figure 3: Macro overview of the analysis pipeline

### 3.2.1. Transformer based Topic Modelling

For the identification of latent themes within research literature, which in its outcomes is similar to traditional topic modelling, we build on top of BERTopic (Grootendorst, 2020).

There are various reasons why we chose a this approach rather than traditional topic modelling techniques such as LDA (Latent Dirichlet Allocation, Blei et al. (2003)) or even simpler algorithms e.g. LSA (Latent Semantic Analysis, Deerwester et al. (1990)). First, we do not have actively preprocess text data, making it easier to transfer this work to other domains and facilitate use by others. While established algorithms are well suited to efficiently handle the size and type of data that we are working with, they require extensive preprocessing of text data before it is passed on. This would typically include, tokenization, removal of stopwords (e.g. and, the, pronouns etc.), taking decisions about potential removal of certain parts of speech (e.g. verbs, prepositions) as well as identification of bi-and n-grams that typically dominate scientific language. Modern transformer language models encapsulate this steps following well documented and benchmarked best practice, thus requiring less expertise within NLP for robust application.

Traditional statistical topic models are typically trained on and applied to the same corpus. That means that they can not "learn" beyond the provided data. In contrast, modern transformer language models are pre-trained on vast amounts of text data and constantly evaluated following transparent protocols. The transfer learning approach, where models trained once on a large amount of data are then fine-tuned and applied to other, smaller data-sets has revolutionised the field of Natural Language Processing particularly since 2018 and the introduction of neural language models like BERT (Devlin et al., 2018). Text representations (embeddings) created using these models are contextualized, capture semantic and syntactic features, handle typical language processing problems such as synonyms and polysemy and have outperformed other approaches in many NLP tasks.

Lastly, we use the hierarchical clustering technique HDBSCAN (McInnes et al., 2017)

for the detection of topics. This allows us to set parameters that directly influence the specificity of identified topics in the literature. Rather than deciding about a fixed amount of topics upfront, we can set a threshold for the minimum number of documents that are allowed to form a topic. Thereby, we can minimize the number of parameters provided by the analyst, making the approach data-driven and easy to replicate/adopt to other domains as possible.

Our approach differs in 2 mayor points from BERTopic (Grootendorst, 2020): We use SPECTER a specialised model for scientific language rather than generic transformer models. We also use NER for the identification of keywords and key-phrases rather than simpler statistical strategies. Figure 4 depicts the different steps of this process.

First, we embed the extracted abstracts using SPECTER (Cohan et al., 2020), a state of the art transformer language model for document-level embedding for scientific text. SPECTER has been developed as a tailor-made embedding model for scientific text and achieved state-of-the-art performance in the majority of document level tasks within the SciDocs evaluation suite. The model is developed departing from SciB-ERT (Beltagy et al., 2019), which in turn has been pre-trained on 1.14M academic papers (full text). While SciBERT is a general language model, developed for NLP tasks within the near textual context (e.g., NER, classification), SPECTER incorporates citation information into the training process to capture signals of inter-document relatedness. This results in text representations that are well suited for "feature-based" downstream tasks, such as clustering or measuring of document similarity (Cohan et al., 2020) utilising other (non-neural) machine learning techniques.

We then use a combination of UMAP (Becht et al., 2019) and HDBSCAN (McInnes et al., 2017) to cluster documents. This combination of dimensionality reduction and density based clustering has shown to scale well while delivering efficient and stable performance, and being able to identify coherent groupings of documents. Discounting for unclustered records and those put into a "catch-all-cluster" we end up with 31957 documents distributed across 218 clusters or topics, with an average size of 148 and a maximum of 863. The minimum size has been set to be 50 as a hyper parameter during the clustering process.

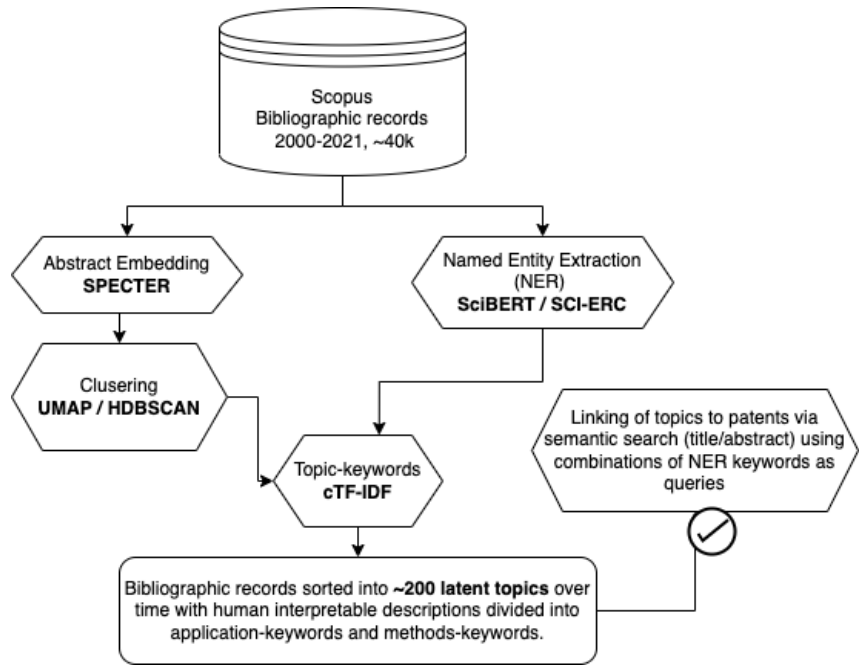

Figure 4: Extended topic modeling pipeline

## Named Entity Extraction for Science-keyword extraction

To create insightful cluster/topic-descriptors we use keywords and key-phrases generated using Named Entity Recognition (NER). Here we utilize a SCIBERT transformer model, which we fine-tune to perform NER. We use the SCIERC dataset (Luan et al., 2018) for the retraining. The dataset consists of 500 abstracts, mainly in the area of computer science, that have been manually annotated to mark *methods, tasks* and *other scientific keywords*. As mentioned above SCIBERT is similarly to SPECTER a large transformer model developed to handle scientific text. Such language models can be retrained to perform specific language tasks such as question answering, summarization or as in our example NER. An advantage of such models is that a limited amount of training data (here, 500 annotated scientific abstracts) are sufficient to achieve good performance. The algorithm identifies and extracts "scientific keywords" sorted into the categories *task, method* and *other scientific terminology* from the abstracts. Despite domain-transfer (from mainly computer science to interdisciplinary neuroscience) the retrained model is able to identify relevant keywords and phrases that can be used to describe detected topics.

We argue that this constitutes a significant improvement over simple frequency-based methods for keyword extraction such as word/n-gram frequencies or algorithms like RAKE (Rose et al., 2010) that are commonly seen in combination with topic modelling exercises. Finally we weight these keywords by importance for the respective cluster/topic with simple TF-IDF (term frequency - inverse document frequency). A collection of keywords and key-phrases that represents an identified topic - using the above mentioned example - would typically look like:

`'sleep', 'REM sleep', 'memory consolidation', 'spindles', 'sleep spindles', 'SWS', 'slow waves', 'slow-wave sleep', 'slow oscillation'`. Here it is important to note, that phrases such as "slow-wave sleep" have been identified automatically, i.e. without the need to specify n-gram length or other parameters.

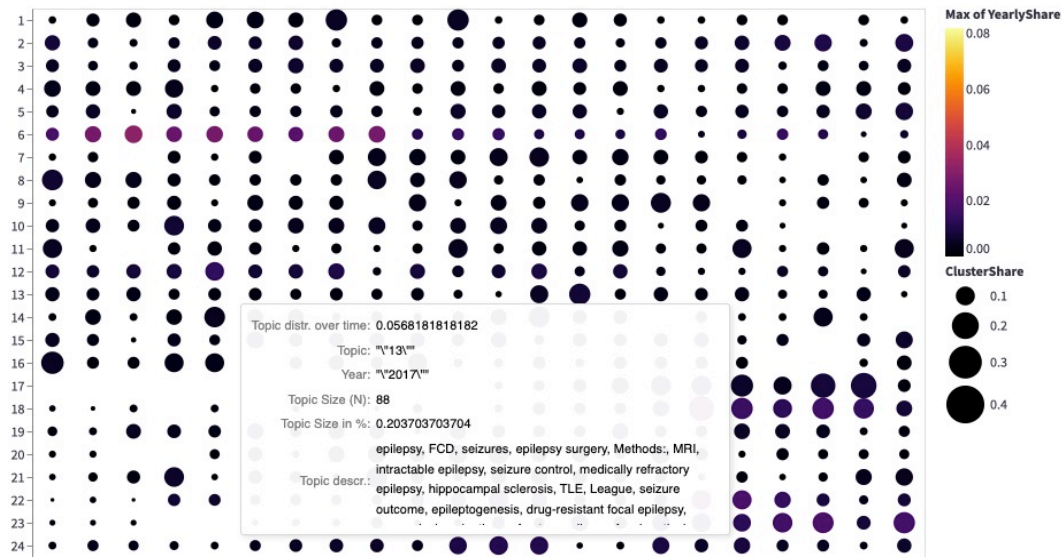

Figure 5: Topics over time (columns) 2000-2021 - screenshot of interactive chart, first 24 topics out of 218

Given that publications have a time dimension (publication date), we can map the development of topics over time and in relation to each other, which is visualized in Figure 5.

Document embeddings are very flexible representations of text. For instance, we can further aggregate these fine-grained topics on higher levels, again using hierarchical clustering on the average-vectors calculated from the individual document vectors belonging to a topic. Figure 6 illustrates how these inter-topic dependencies can be presented in a dendogramm.

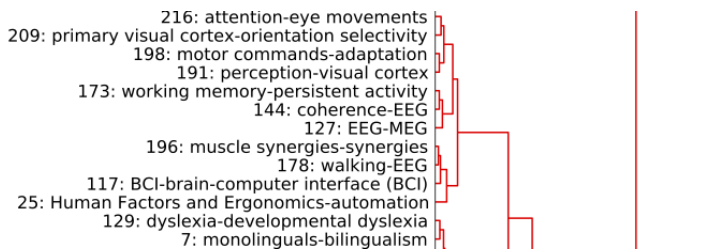

Figure 6: Cropped dendogramm, showing the relationship of individual topics

### 3.2.2. Science-technology linkage

One could argue that the presented approach aims at mimicking a domain expert that is performing a technology screening exercise to scope the patent space e.g. to assess the novelty for an invention. In fact, that is not far from the reality, where patent experts identify relevant keywords and hand-craft queries within specific technology areas. Our approach operates similarly: For each of the earlier identified topics, we use the top100 keywords (within the 2 domains of methods and applications) to generate queries. These are 50 random 25-keywords-long sequences combining evenly both types of keywords. To perform "free-text-semantic-search" on the whole PATSTAT corpus, we use the embedding and semantic search methodology proposed in Hain et al. (2022a). Here, abstracts are embedded using a custom-trained Word2Vec (Mikolov et al., 2013) and TF-IDF model, resulting in dense TF-IDF weighted embeddings. Then, nearest neighbor approximation is used, utilizing the efficient Annoy (Bernhardsson, 2017) approach, to identify closest matches.

Given that the search strings generated from scientific literature are keyword based, i.e. the queries do not have an explicit sequence that carries some meaning, we can transform them into the same vector space as the patent abstracts. We then perform 10.090 searches, asking the system to return 100 closest patents to the search query.

## 4. Analysis

In this section we present overall results of the exercise, i.e. summaries for all topics, attributed publications and related patents. We then zoom in on two topics that undoubtedly can be linked to neurotechnology – (1) Brain Comptuer Interfaces (BCI) and (2) Deep Brain Stimulation (DBS). Note that this version represents an early draft including only preliminary results. [6]

### 4.1. Neuroscience topics in scientific publications

The topic modelling approach described above identifies 218 fine grained topics. We opted for a larger number of identified topics because it is much easier to join two of them in the further process if needed than having to subdivide one larger topic if it is deemed to lack nuance. Future versions of this research will include an expert evaluation at this stage to validate relevance and specificity of identified topics.

As depicted in Figure 5 one can see that topics are comprised of different numbers of articles (in the overall corpus). The number changes over time and can be interpreted as a proxy for relevance of a topic over time and relative to other topics in the respective year.

An example of the functionality that does not require much expertise is presented in Figure 7. Here we see 2 (out of several) COVID-19 related topics that appear in 2020 and become dominant for the years 2020 and 2021. Here our approach is able to detect an emerging discussion distinguishing between discipline-specific perspectives on the pandemic.

Similarly we observe the gradual growth of topics related to machine learning and artificial intelligence, covering work on, e.g., feature extraction from EEG signals for automated epilepsy seizure detection. Other areas such as work that links attention with eye movements and the visual cortex seem to lose importance over time. This seems to align well with over-time publication numbers found on Scopus, see Fig-ure A.5. Overall, we can conclude that the presented approach identifies nuanced and useful groupings of literature. It also extracts keywords that provide specific and clear descriptions. However, the results require expert validation to confirm these claims as well as to ascertain that the identified keywords, in fact provide a solid foundation to conduct patent searches in different areas.

---

[6]Results can be explored in detail in an interactive dashboard app that summarizes the results of our analysis https://neurscience-sci-pat.herokuapp.com/

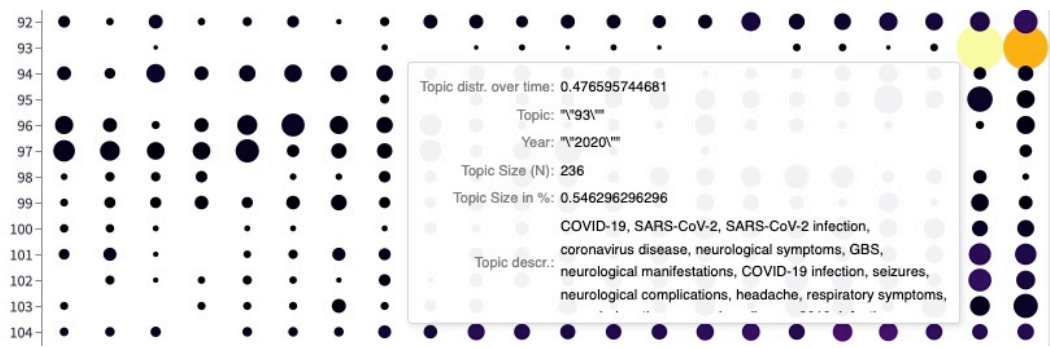

(a) COVID-19 and Neurological Symptoms

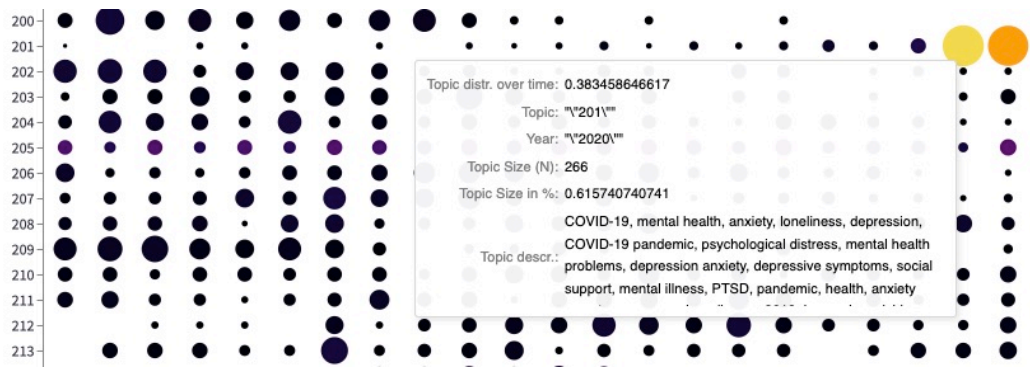

(b) COVID-19 effects on Mental Health

Figure 7: COVID-19 related publications within different nuanced topics

## 4.2. Neuroscience technologies in patents

Using our dataset of IP5 priority patents, we query their abstract texts via the semantic search of search described in section 3, and obtain 219.501 unique patent applications as results. In the following we provide a preliminary overview over the characteristics and distribution of identified neurotechnology patents around the globe.

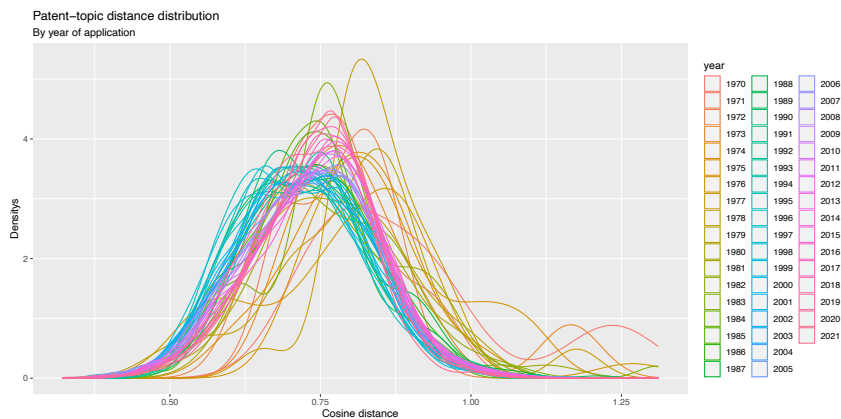

Figure 8: Patent-topic distance distribution by year

We also capture a semantic cosine distance score that denotes a ranking in terms of closeness of the identified patent to the keyword-query, and the associated technology topic. Figure 8 illustrates the distribution of cosine distance by year. We generally observe this distance to be normally distributed, yet also observe variation in distance across the years. This might indicate–and will be used to analyze–periods of more or less rapid technological change, as well as times of changing techno-economic paradigms (Perez, 2010). Figure 9 illustrates the popularity of the top technology clusters, highlighting that all most popular technologies have experienced a sharp increase post 2015.

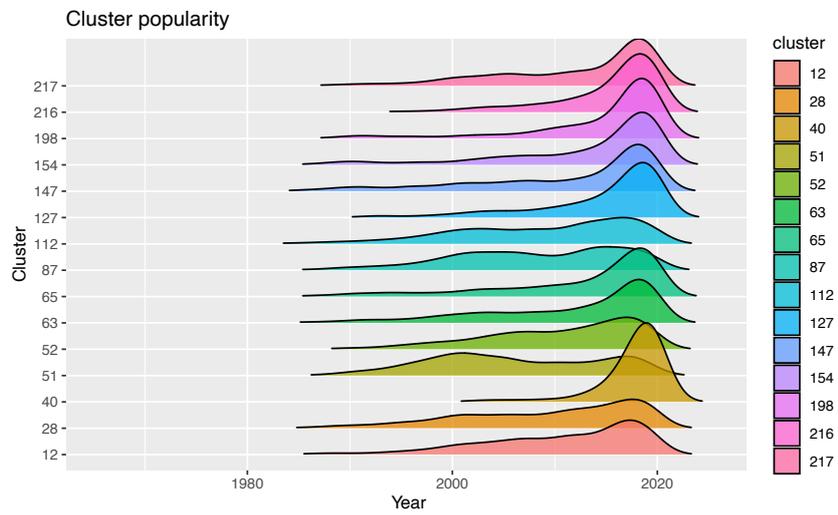

Figure 9: Neuroscience patent s by cluster

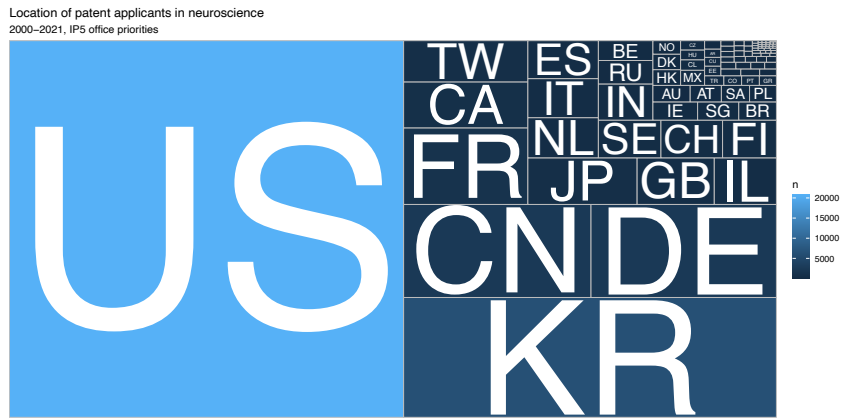

Figure 10: Top neuroscience countries by applicant

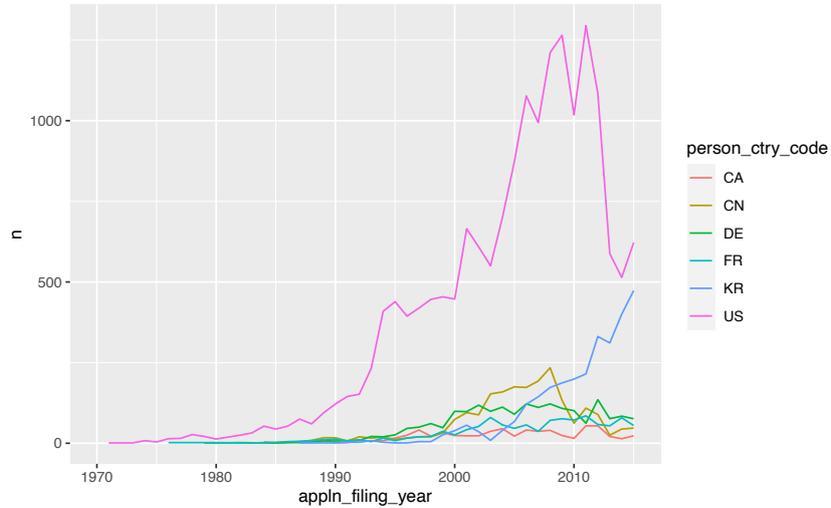

Figure 11: Neuroscience patent application by year and top country

Figure 10 illustrates the location of applicants of identified neurotechnology patents. We see a similar composition as in neuroscience publications (cf. 1), including the US dominance. The only major difference in terms of country ranking is the more pronounced position of South Korea in neurotechnology applications as compared to scientific production. Overall, the results overlap with established insights on scioence-technology linkages, and the interaction between the local scientific knowledge base with technology development (e.g., at the case of artificial intelligence technologies, cf. Klinger et al., 2021; Baruffaldi et al., 2020). Figure 11 shows this development over time. We see the dominant position of the United States since 2010 to be in decline, while South Korea is rapidly catching up.

In terms of technological composition of associated neurotechnology patents, a breakdown of their technology fields can be found in Figure 12. Interestingly, Computer Technology appears to be the most significant technology field, ahead of more expected fields such as Medical Technology, Biotechnology, and Pharmaceuticals. This hints at the growing importance of algorithmic applications in general, including neutral computing techniques which are rapidly increasing in popularity and impact.

Figure 13 displays the relationship between technology fields of neurotechnologies in a technology space network, where the nodes represent technology fields, and edges their relatedness (Balland et al., 2019). We here see three main clusters around i.) chemistry related fields, ii.) digital technologies, and iii.) engineering.

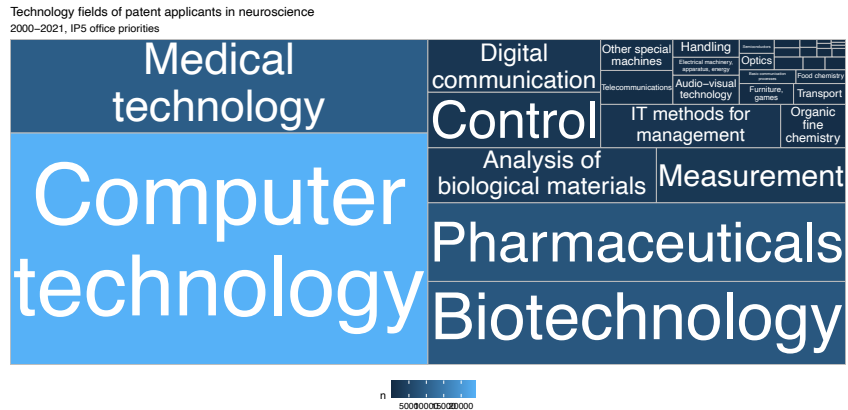

Figure 12: Top neuroscience technology fields

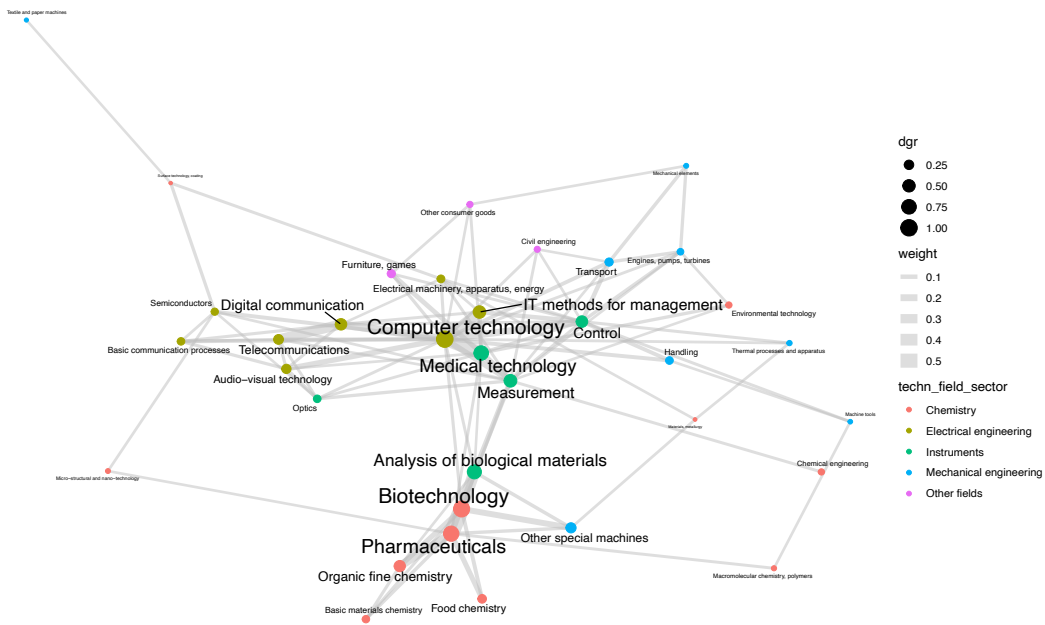

Figure 13: Relatedness of neuroscience technology fields

### 4.3. Brain Computer Interfaces (BCI)

In the scientific literature we identify 212 publications that can be attributed to this topic. The number of publications peaks in 2010 remaining relatively stable. Articles outline different invasive and non-invasive within this domain, including recently application of machine learning techniques to support BCI, but we find also work looking into ethical challenges e.g., "An Analysis of the Impact of Brain-Computer Interfaces on Autonomy".

Among the matched patents, we identify technologies at the intersection of medical equipment, measuring devices, computing and electronic communication. Where it comes to disclosed applications many of these seem to be related to rehabilitation and support for disabled patients (e.g., brain-based spelling method, wheelchair control devices, respiratory training systems). Most patents can be attributed to USA, China and UK, also when the semantic distance threshold.

### 4.4. Deep Brain Stimulation (DBS)

In the scientific literature we identify 169 publications in this domain in our corpus. It is a growing topic over time, especially since 2015.
Articles describe methods where DBS is used to treat neurological conditions but also mental illnesses such as obsessive-compulsive disorder (OCD). Similarly to the case of BCI, we identify many related patents that describe technologies with similar combinations of patent classes. Patent documents describe techniques using magnetic and electrical stimulation of the brain and different parts of the nervous system for therapeutic purposes. The amount of identified related patents is smaller as for BCI, with the USA, China, Japan and South Korea leading the ranking.

## 5. Conclusion

In this paper, we presented an efficient deep learning based approach to extract technology-related topics and keywords within scientific literature, and identify corre-sponding technologies within patent applications. We illustrated the workflow as well as results obtained by mapping publications within the field of neuroscience to related patent applications, aiming at the mapping of neurotechnology, and particularly the identification of emerging ones.

Specifically, we utilize transformer based language models, tailored for use with scientific text, to detect coherent topics over time and describe these by relevant keywords that are automatically extracted from a large text corpus. We identify these keywords using Named Entity Recognition, distinguishing between those describing methods, applications and other scientific terminology. These topics are created via density-based clustering of transformer based embeddings, which are fine-tuned to scientific literature. In contrast to traditional topic modelling techniques, our approach produces topics focused on the description of technologies and their applications rather than general themes in the corpus. We create a large amount of search queries based on combinations of method- and application-keywords, which we use to conduct semantic search and identify related patents.

By doing so, we aim at contributing to the growing body of research on text-based technology mapping and forecasting that leverages latest advances in natural language processing and deep learning. We demonstrate at the case of neuroscience research, that the developed approach is able to extract technology topics in broad, interdisciplinary, and dynamic field, and map these topics to patent data. Enabling the semi-automatized mapping of technologies identified in scientific literature to patent applications, we are thereby providing an empirical foundation for the study of science-technology linkages.

The presented method as well as the obtained preliminary results are at its current stage subject to a number of limitations. As discussed in section 3, our main aim is to detect technology topics in scientific publications, use them to identify science-technology linkages, and finally map the development of neurotechnologies in patent data. To do so we create a neuroscience related corpus of scientific literature by filtering Scopus by subject area. Subject areas are assigned on journal level, and rather broad, in this case ranging from computer science, over chemistry, biology, and psychology. While this is an inherent feature of neuroscience research, it complicates the identification of technology topics from the publications text data. While our approach is geared towards selecting terms related to scientific methods, many of them cannot be related to an actual technology, but rather a method of scientific inquiry within academia. A more focused and technology-targeted selection of publications could limit this effect. However, within the presented approach, the manual selection of relevant technology topics by a domain expert remains necessary.

Furthermore, the presented approach assumes technologies to initially emerge in scientific literature, and then later being further developed to commercial technologies. Yet, it is reasonable that distinct technologies are emerging without former traces in science, or in commercial applications develop far away from their scientific origin in terms of methods and techniques as well as issues addressed. This could to some extent be addressed by extending the approach presented in this paper with additional iterative steps to identify additional technology topics within the selected patents, and expanding this selection to similar patents outside the current selection.

# References

Acosta, M. and Coronado, D. (2003). Science–technology flows in spanish regions: an analysis of scientific citations in patents. *Research Policy*, 32(10):1783–1803.

Adner, R. and Kapoor, R. (2010). Value creation in innovation ecosystems: How the structure of technological interdependence affects firm performance in new technology generations. *Strategic management journal*, 31(3):306–333.

Allhoff, F., Lin, P., and Steinberg, J. (2011). Ethics of human enhancement: an executive summary. *Science and engineering ethics*, 17(2):201–212.

Anderson, J., Mizgalewicz, A., and Illes, J. (2012). Reviews of functional mri: The ethical dimensions of methodological critique.

Bakhtin, P., Saritas, O., Chulok, A., Kuzminov, I., and Timofeev, A. (2017). Trend monitoring for linking science and strategy. *Scientometrics*, 111(3):2059–2075.

Balland, P.-A., Boschma, R., Crespo, J., and Rigby, D. L. (2019). Smart specialization policy in the european union: relatedness, knowledge complexity and regional diversification. *Regional studies*, 53(9):1252–1268.

Baruffaldi, S., van Beuzekom, B., Dernis, H., Harhoff, D., Rao, N., Rosenfeld, D., and Squicciarini, M. (2020). Identifying and measuring developments in artificial intelligence: Making the impossible possible.

Becht, E., McInnes, L., Healy, J., Dutertre, C.-A., Kwok, I. W., Ng, L. G., Ginhoux, F., and Newell, E. W. (2019). Dimensionality reduction for visualizing single-cell data using umap. *Nature biotechnology*, 37(1):38–44.

Bekkers, R. and Freitas, I. M. B. (2008). Analysing knowledge transfer channels between universities and industry: To what degree do sectors also matter? *Research policy*, 37(10):1837–1853.

Beltagy, I., Lo, K., and Cohan, A. (2019). Scibert: Pretrained language model for scientific text. In *EMNLP*.

Bernhardsson, E. (2017). Annoy: approximate nearest neighbors in c++/python optimized for memory usage and loading/saving to disk. *GitHub https://github.com/spotify/annoy*.

Blei, D. M., Ng, A. Y., and Jordan, M. I. (2003). Latent dirichlet allocation. *Journal of machine Learning research*, 3(Jan):993–1022.

Boyack, K. W. and Klavans, R. (2008). Measuring science–technology interaction using rare inventor–author names. *Journal of Informetrics*, 2(3):173–182.

Butler, D. (2016). Tomorrow's world: technological change is accelerating today at an unprecedented speed and could create a world we can barely begin to imagine. *Nature*, 530(7591):398–402.

Callaert, J., Grouwels, J., and Van Looy, B. (2012). Delineating the scientific footprint in technology: Identifying scientific publications within non-patent references. *Scientometrics*, 91(2):383–398.

Callaert, J., Van Looy, B., Verbeek, A., Debackere, K., and Thijs, B. (2006). Traces of prior art: An analysis of non-patent references found in patent documents. *Scientometrics*, 69(1):3–20.

Cassiman, B., Glenisson, P., and Van Looy, B. (2007). Measuring industry-science links through inventor-author relations: A profiling methodology. *Scientometrics*, 70(2):379–391.

Catalán, P., Navarrete, C., and Figueroa, F. (2020). The scientific and technological cross-space: is technological diversification driven by scientific endogenous capacity? *Research Policy*, page 104016.

Cohan, A., Feldman, S., Beltagy, I., Downey, D., and Weld, D. S. (2020). SPECTER: Document-level Representation Learning using Citation-informed Transformers. In *ACL*.

De Rassenfosse, G., Dernis, H., Guellec, D., Picci, L., and de la Potterie, B. v. P. (2013). The worldwide count of priority patents: A new indicator of inventive activity. *Research Policy*, 42(3):720–737.

De Rassenfosse, G., Kozak, J., and Seliger, F. (2019). Geocoding of worldwide patent data. *Nature Scientific Data*, 6(1):1–15.

Deerwester, S., Dumais, S. T., Furnas, G. W., Landauer, T. K., and Harshman, R. (1990). Indexing by latent semantic analysis. *Journal of the American society for information science*, 41(6):391–407.

Devlin, J., Chang, M.-W., Lee, K., and Toutanova, K. (2018). Bert: Pre-training of deep bidirectional transformers for language understanding. *arXiv preprint arXiv:1810.04805*.

Drew, L. (2019). The ethics of brain-computer interfaces. *Nature*, 571(7766):S19–S19.

Dubljević, V., Jox, R. J., and Racine, E. (2017). Neuroethics: Neuroscience's contributions to bioethics. *Bioethics*, 31(5).

Etzkowitz, H. and Leydesdorff, L. (2000). The dynamics of innovation: from national systems and "mode 2" to a triple helix of university–industry–government relations. *Research policy*, 29(2):109–123.

Garden, H., Winickoff, D. E., Frahm, N. M., and Pfotenhauer, S. (2019). Responsible innovation in neurotechnology enterprises.

Giunta, A., Pericoli, F. M., and Pierucci, E. (2016). University–industry collaboration in the biopharmaceuticals: The italian case. *The Journal of Technology Transfer*, 41(4):818–840.

Glänzel, W. and Meyer, M. (2003). Patents cited in the scientific literature: An exploratory study of'reverse'citation relations. *Scientometrics*, 58(2):415–428.

Grootendorst, M. (2020). Bertopic: Leveraging bert and c-tf-idf to create easily interpretable topics.

Hain, D. S., Jurowetzki, R., Buchmann, T., and Wolf, P. (2022a). A text-embedding-based approach to measuring patent-to-patent technological similarity. *Technological Forecasting and Social Change*, 177:121559.

Hain, D. S., Jurowetzki, R., Zhou, Y., and Lee, S. (2022b). Introduction to the special issue: Machine learning and ai for science, technology, and (eco-)system mapping and forecasting. *Scientometrics*, (forthcoming).

Henderson, R., Jaffe, A., and Trajtenberg, M. (1994). Numbers up, quality down? trends in university patenting 1965-1992, presentation to cepr/aaas conference university goals, institutional mechanisms and the industrial transferability of research.

Jarchum, I. (2019). The ethics of neurotechnology. *Nature Biotechnology*, 37(9):993–996.

Klinger, J., Mateos-Garcia, J., and Stathoulopoulos, K. (2021). Deep learning, deep change? mapping the evolution and geography of a general purpose technology. *Scientometrics*, 126(7):5589–5621.

Looy, B. V., Zimmermann, E., Veugelers, R., Verbeek, A., Mello, J., and Debackere, K. (2003). Do science-technology interactions pay off when developing technology? *Scientometrics*, 57(3):355–367.

Luan, Y., He, L., Ostendorf, M., and Hajishirzi, H. (2018). Multi-task identification of entities, relations, and coreferencefor scientific knowledge graph construction. In *Proc. Conf. Empirical Methods Natural Language Process. (EMNLP)*.

Lundvall, B. (1992). *National Systems of Innovation: Toward a Theory of Innovation and Interactive Learning*. Anthem Other Canon Economics. Anthem Press.

Magerman, T., Van Looy, B., and Debackere, K. (2015). Does involvement in patenting jeopardize one's academic footprint? an analysis of patent-paper pairs in biotechnology. *Research Policy*, 44(9):1702–1713.

Mansfield, E. (1991). Academic research and industrial innovation. *Research policy*, 20(1):1–12.

McInnes, L., Healy, J., and Astels, S. (2017). hdbscan: Hierarchical density based clustering. *Journal of Open Source Software*, 2(11):205.

McMillan, G. S., Narin, F., and Deeds, D. L. (2000). An analysis of the critical role of public science in innovation: the case of biotechnology. *Research policy*, 29(1):1–8.

Meyer, M. (2000). Does science push technology? patents citing scientific literature. *Research Policy*, 29(3):409–434.

Meyer-Krahmer, F. and Schmoch, U. (1998). Science-based technologies: university–industry interactions in four fields. *Research policy*, 27(8):835–851.

Michel, J. and Bettels, B. (2001). Patent citation analysis. a closer look at the basic input data from patent search reports. *Scientometrics*, 51(1):185–201.

Mikolov, T., Sutskever, I., Chen, K., Corrado, G. S., and Dean, J. (2013). Distributed representations of words and phrases and their compositionality. In Burges, C. J. C., Bottou, L., Welling, M., Ghahramani, Z., and Weinberger, K. Q., editors, *Advances in Neural Information Processing Systems*, volume 26. Curran Associates, Inc.

Narin, F., Hamilton, K. S., and Olivastro, D. (1997). The increasing linkage between us technology and public science. *Research policy*, 26(3):317–330.

Nelson, R. R. (1993). *National Innovation Systems: A Comparative Analysis*. Oxford University Press.

Noyons, E., van Raan, A., Grupp, H., and Schmoch, U. (1994). Exploring the science and technology interface: inventor-author relations in laser medicine research. *Research Policy*, 23(4):443–457.

Park, Y. and Yoon, J. (2017). Application technology opportunity discovery from technology portfolios: Use of patent classification and collaborative filtering. *Technological Forecasting and Social Change*, 118:170–183.

Perez, C. (2010). Technological revolutions and techno-economic paradigms. *Cambridge journal of economics*, 34(1):185–202.

Porter, A. and Rafols, I. (2009). Is science becoming more interdisciplinary? measuring and mapping six research fields over time. *Scientometrics*, 81(3):719–745.

Porter, A. L. and Cunningham, S. W. (2004). *Tech mining: Exploiting new technologies for competitive advantage*. John Wiley & Sons.

Ranaei, S., Suominen, A., and Dedehayir, O. (2017). A topic model analysis of science and technology linkages: A case study in pharmaceutical industry. In *2017 IEEE Technology & Engineering Management Conference (TEMSCON)*, pages 49–54. IEEE.

Rose, S., Engel, D., Cramer, N., and Cowley, W. (2010). Automatic keyword extraction from individual documents. *Text mining: applications and theory*, 1:1–20.

Savage, N. (2019). How ai and neuroscience drive each other forwards. *Nature*, 571(7766):S15–S15.

Schmoch, U. (1997). Indicators and the relations between science and technology. *Scientometrics*, 38(1):103–116.

Schwechheimer, H. and Winterhager, M. (2001). Mapping interdisciplinary research fronts in neuroscience: A bibliometric view to retrograde amnesia. *Scientometrics*, 51(1):311–318.

Shibata, N., Kajikawa, Y., and Sakata, I. (2011). Detecting potential technological fronts by comparing scientific papers and patents. *Foresight*.

Soo, V.-W., Lin, S.-Y., Yang, S.-Y., Lin, S.-N., and Cheng, S.-L. (2006). A cooperative multi-agent platform for invention based on patent document analysis and ontology. *Expert Systems with Applications*, 31(4):766–775.

Squicciarini, M., Dernis, H., and Criscuolo, C. (2013). Measuring patent quality: Indicators of technological and economic value.

Tussen, R., Buter, R., and Van Leeuwen, T. (2000). Technological relevance of science: An assessment of citation linkages between patents and research papers. *Scientometrics*, 47(2):389–412.

Verbeek, A., Debackere, K., Luwel, M., Andries, P., Zimmermann, E., and Deleus, F. (2002). Linking science to technology: Using bibliographic references in patents to build linkage schemes. *Scientometrics*, 54(3):399–420.

Yang, C., Zhu, D., Wang, X., Zhang, Y., Zhang, G., and Lu, J. (2017). Requirement-oriented core technological components' identification based on sao analysis. *Scientometrics*, 112(3):1229–1248.

Yeung, A. W. K., Goto, T. K., and Leung, W. K. (2017). The changing landscape of neuroscience research, 2006–2015: a bibliometric study. *Frontiers in neuroscience*, 11:120.

Zhang, Y., Zhang, G., Chen, H., Porter, A. L., Zhu, D., and Lu, J. (2016). Topic analysis and forecasting for science, technology and innovation: Methodology with a case study focusing on big data research. *Technological forecasting and social change*, 105:179–191.

# A. Appendix

## A.1. Neuroscience Publications

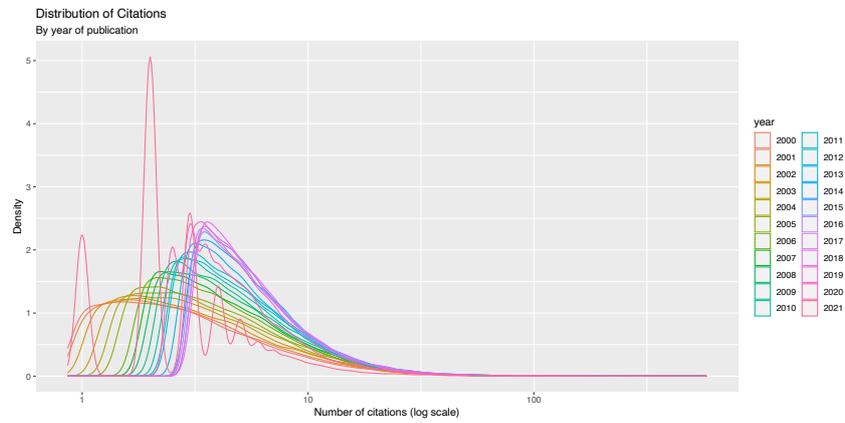

Figure A.1: Citation distribution by year of publication

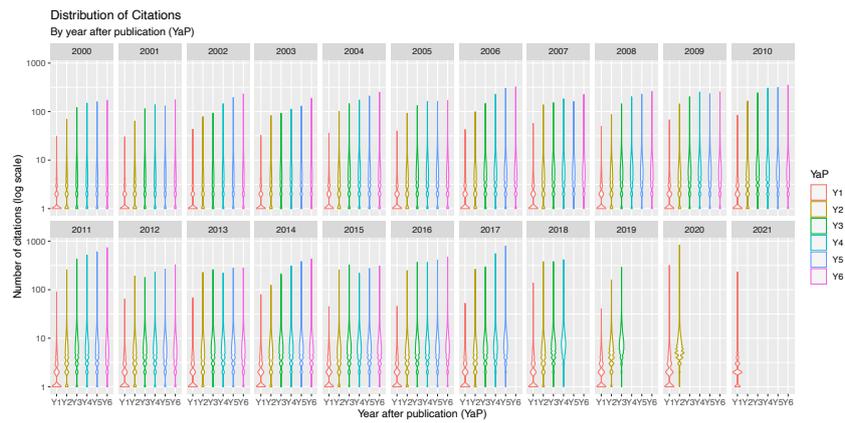

Figure A.2: Citation distribution by year after publication

## A.2. Neuroscience Patents

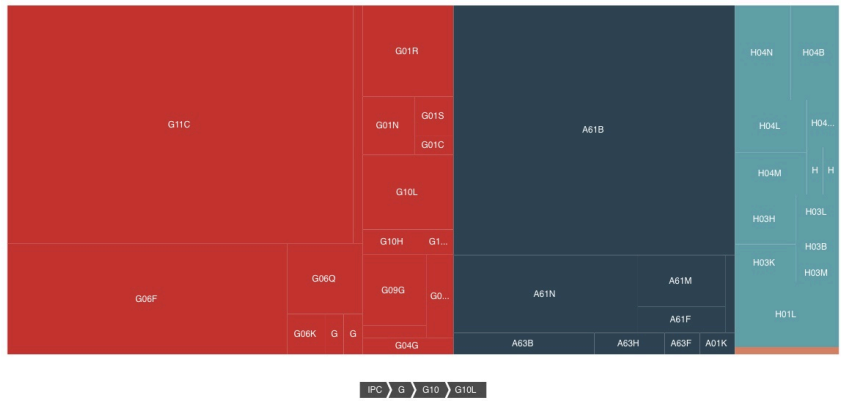

Figure A.3: Overview of patent classes for the "Sleep and Memory Consolidation" topic

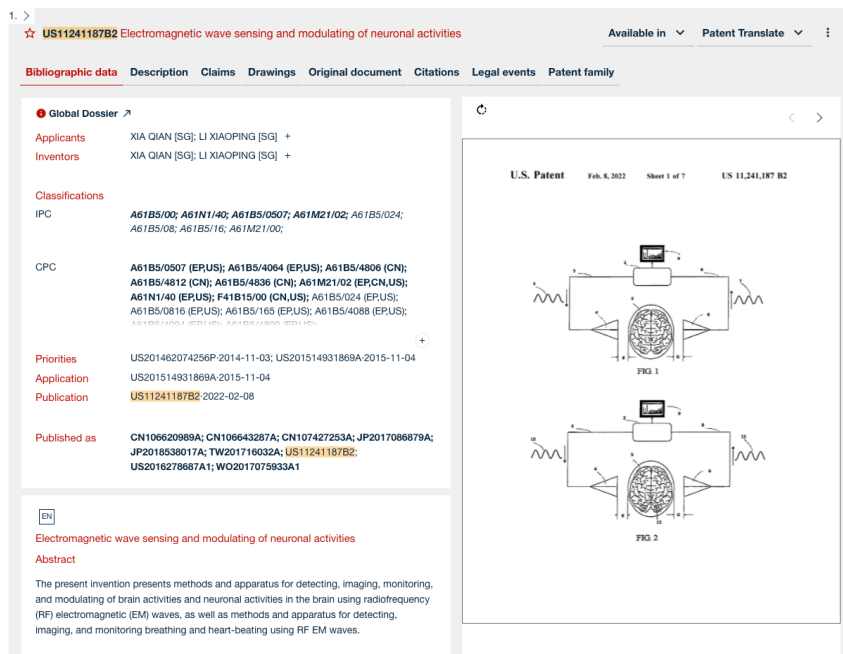

Figure A.4: Patent document with low semantic distance belonging to the "Sleep and Memory Consolidation" topic

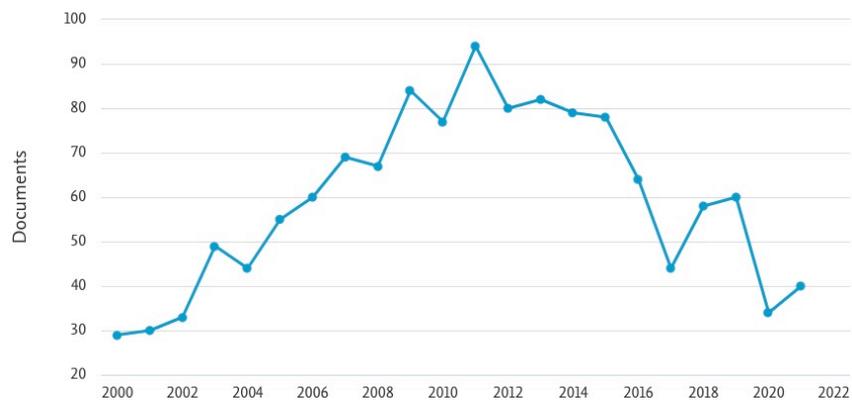

Figure A.5: Scopus documents over time for search-string: attention AND eye AND movements AND visual AND cortex